\documentclass[aps,twocolumn,superscriptaddress,nofootinbib,a4paper,longbibliography]{revtex4-2}

\newcommand{\stt}{\mathrm{s.t.}}
\newcommand{\E}{\mathcal{E}}
\usepackage{times}
\usepackage{epsfig}
\usepackage{amsfonts}
\usepackage{amsmath}
\usepackage{amsthm}
\usepackage{amssymb}					
\usepackage{amsthm}
\usepackage{dsfont}
\usepackage{bm}
\usepackage{mathtools}
\usepackage{color}
\usepackage{multirow}
\usepackage[normalem]{ulem}
\newcommand{\stkout}[1]{\ifmmode\text{\sout{\ensuremath{#1}}}\else\sout{#1}\fi}
\usepackage{latexsym}
\usepackage{mathrsfs}
\usepackage{natbib}
\usepackage{verbatim}
\usepackage[T1]{fontenc}
\usepackage{float}
\usepackage{graphicx}
\usepackage{xcolor}
\usepackage{physics}
\usepackage{soul}
\usepackage{subcaption}

\usepackage[font=small,labelfont=bf]{caption}

\newtheorem{theorem}{Theorem}

\newtheorem{observation}[]{Observation}
\newtheorem{result}[]{Result}

\newtheorem{definition}[theorem]{Definition}

\usepackage[colorlinks=true,linkcolor=blue,citecolor=magenta,urlcolor=blue]{hyperref}

\begin{document}
	

	\title{Absolute dimensionality of quantum ensembles}

	\author{Alexander Bernal}
	\affiliation{Instituto de F\'isica Te\'orica, IFT-UAM/CSIC, 	Universidad Aut\'onoma de Madrid,
		Cantoblanco, 28049 Madrid, Spain}	
	
	\author{Gabriele Cobucci}
	\affiliation{Physics Department and NanoLund, Lund University, Box 118, 22100 Lund, Sweden.}
	
	\author{Martin J. Renner}
	\affiliation{University of Vienna, Faculty of Physics and VDSP, Vienna Center for Quantum Science and Technology (VCQ), Boltzmanngasse 5, 1090 Vienna, Austria.}	
	\affiliation{Institute for Quantum Optics and Quantum Information (IQOQI), 	Austrian Academy of Sciences, Boltzmanngasse 3, 1090 Vienna, Austria}

	\author{Armin Tavakoli}
	\affiliation{Physics Department and NanoLund, Lund University, Box 118, 22100 Lund, Sweden.}

	\begin{abstract}
		The dimension of a quantum state is traditionally seen as the number of superposed distinguishable states in a given basis. We propose an absolute, i.e.~basis-independent, notion of dimensionality for ensembles of quantum states. It is based on whether a quantum ensemble  can be simulated with states confined to arbitrary lower-dimensional subspaces and classical postprocessing. In order to determine the absolute dimension of quantum ensembles, we develop both analytical witness criteria and a semidefinite programming criterion based on the ensemble's information capacity. Furthermore, we construct explicit simulation models for arbitrary ensembles of pure quantum states subject to white noise, and in natural cases we prove their optimality. Also, efficient numerical methods are provided for simulating generic ensembles. Finally, we  discuss the role of absolute dimensionality in high-dimensional quantum information processing.
	\end{abstract}
	
	\date{\today}

	\maketitle
	
	\textit{Introduction.---} Hilbert space dimension is an essential property of quantum systems. It reflects the assumptions made about a system, corresponding to the number of independent degrees of freedom under control. For instance, it is central for the definition of the qubit and natural for comparing classical and quantum states. Therefore, there has been much interest  in studying the dimensionality of various aspects of quantum operations, for example entanglement \cite{Terhal2000}, dynamics \cite{Wolf2009}, measurements \cite{Ioannou2022} and Bell nonlocality \cite{Navascues2014}. This parallels the emerging importance of high-dimensional systems in quantum technology, in e.g.~communication \cite{Cozzolino2019}, entanglement distribution \cite{Erhard2020} and computation \cite{Wang2020}.

	Traditionally, the dimension, $d$, of a pure quantum state $\ket{\psi}=\sum_{k=1}^dc_k\ket{e_k}$, with $c_k\in\mathbb{C}$, is the number of distinguishable states appearing in the superposition. This supposes the existence of a priviledged orthonormal basis, $\{\ket{e_k}\}_k$, which constitutes the reference frame with respect to which the superposition is defined. This can be more or less natural depending on the specific system, but it ultimately restricts the generality of the dimension concept and its potential applications. 
	
	For instance, while extremal sets of pure $d$-dimensional states can be self-tested \cite{Tavakoli2018, Navascues2023}, a small amount of noise is usually sufficient to rapidly weaken the certification \cite{Farkas2019, Rosset2019}. This makes it non-trivial to determine if even  $d$-dimensional states are necessary, but addressing this requires a basis-independent approach to dimensionality. Complementary to this is the well-known notion of device-independent dimension witnessing \cite{Gallego2010, Ahrens2012, Hendrych2012}. In this task, one is interested in bounding the dimension of a quantum ensemble but resorts to inferring the bound only from the input/output statistics it produces. That the ensemble admits a $d$-dimensional realisation is a necessary but not sufficient condition for it being detectable as such on the level of correlations. In order to meaningfully  address the ensemble dimension directly, on the level of Hilbert space formalism, one needs a basis-independent notion  of dimensionality. 
	
	\begin{figure}[t!]
		\centering
		\includegraphics[width=1\columnwidth]{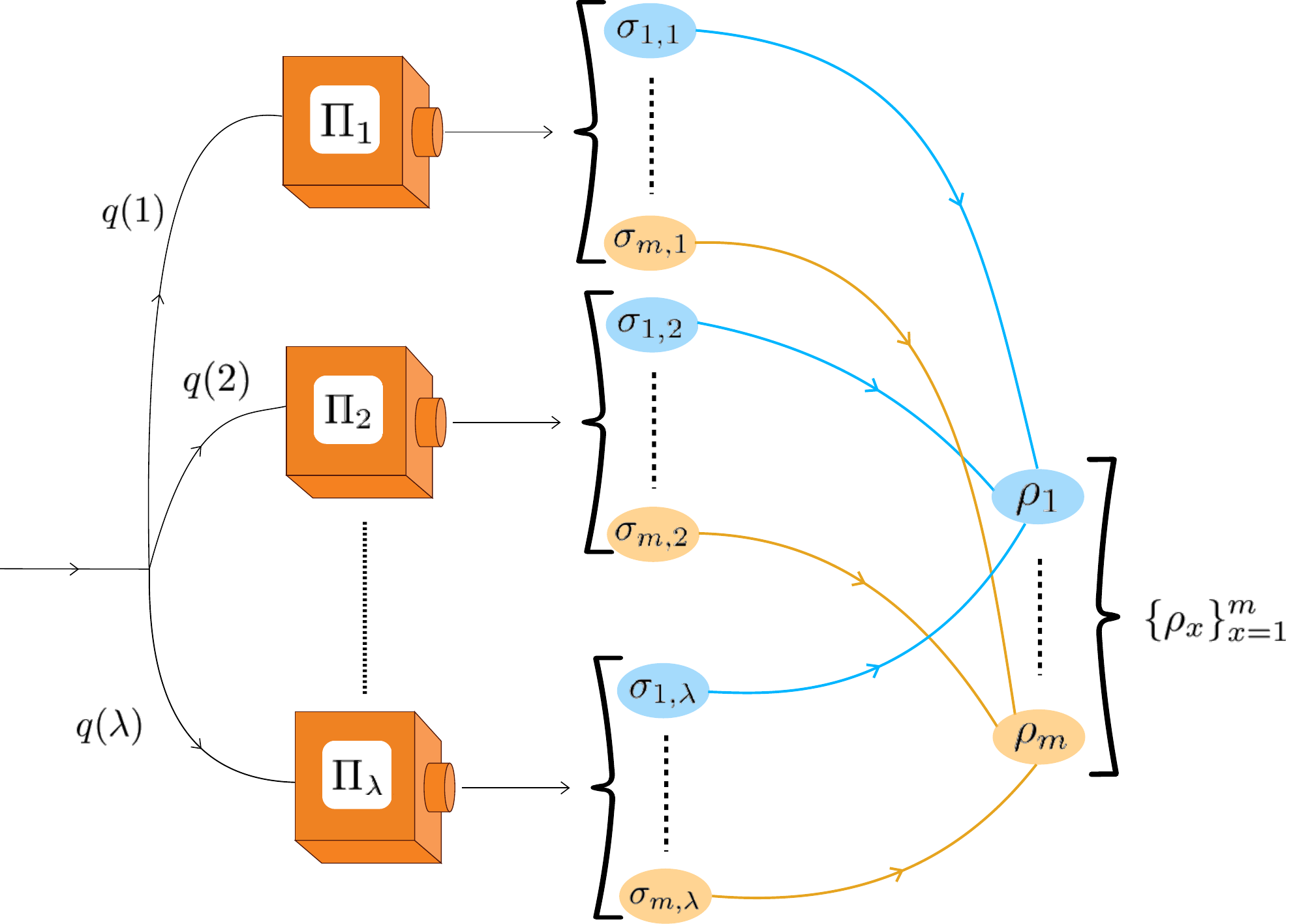}
		\caption{\textbf{Simulation of ensemble.} Given a $d$-dimensional ensemble $\mathcal{E}=\{\rho_x\}_{x=1}^m$ of $m$ states, can it be simulated by a set of quantum preparation devices that are, respectively, restricted to $r$-dimensional subspaces  $\lbrace \Pi_\lambda \rbrace$? Each device is used with probability $q_\lambda$ and the prepared states $\sigma_{x,\lambda}$ can then be classically postprocessed. The  absolute dimension of $\mathcal{E}$ is the smallest $r$ for which a simulation exists.}
		\label{Fig_simulation_def1}
	\end{figure}

	Here, we propose a notion of absolute dimensionality for ensembles of quantum states. It is basis-independent and operationally intuitive.
	The main idea is whether a set of preparation devices that are each restricted to some $r$-dimensional subspace, with $r< d$, are sufficient to reconstruct the original, $d$-dimensional, quantum ensemble. This is illustrated in Fig.~\ref{Fig_simulation_def1}. If affirmative, we conclude that lower-dimensional quantum states can substitute for the original quantum ensemble, and the smallest $r$ for which this is possible becomes the absolute dimension of the ensemble. We first formalise this concept and then develop criteria for characterising it for arbitrary quantum ensembles.  We present two general and practical methods for witnessing absolute dimensionality. The first is based on bounding the value of suitable observables, whose violation implies a lower bound on the absolute dimension. Importantly, we show that these criteria can often be solved exactly. The second is based on identifying a simple connection to quantum state discrimination, inspired by the concept of  informationally-restricted quantum correlations \cite{Tavakoli2020informationally, Tavakoli2022informationally}. This yields a semidefinite  programming \cite{tavakoli2024semidefiniteprogrammingrelaxationsquantum} criterion for the absolute dimension. Furthermore, we then take also the opposite perspective and ask whether explicit lower-dimensional simulations can be constructed for high-dimensional quantum states. We put forward a model that can simulate any number of pure states subject to a critical rate of isotropic noise, and we prove its optimality for large classes of ensembles. In order to address also generic types of noise, we develop an efficient numerical method for constructing simulation and we showcase its advantages in simple case studies.

	\textit{Quantum absolute dimension.---} Consider a set of $m$ quantum states $\mathcal{E} = \lbrace \rho_x \rbrace_{x = 1}^{m}$, where all states are defined on a $d$-dimensional Hilbert space. In order to characterise the dimensionality of this set without referring to any specific basis, we ask whether it is possible to simulate  $\mathcal{E}$ by using only $r$-dimensional quantum systems, with $r<d$, and classical randomness and postprocessing. To this end, consider a procedure in which we have access to a potentially unboundedly large set of preparation devices, $\lbrace \mathcal{P}_{\lambda} \rbrace$, each of which can emit arbitrary states confined to an $r$-dimensional subspace corresponding to the rank-$r$ projector $\Pi_\lambda$. Importantly, these subspaces can be arbitrary too, only their dimension is restricted.  The preparations outputted by $\mathcal{P}_{\lambda}$ are denoted  $\lbrace \tau_{z,\lambda} \rbrace_{z}$, and the probability density of consulting $\mathcal{P}_\lambda$ is denoted $q(\lambda)$. Then, we aim to reconstruct $\mathcal{E}$ via a set of classical postprocessing rules $\{p(z|x,\lambda)\}$. Thus, the simulation takes the form
	\begin{equation}
	\rho_x=\int d\lambda \ q(\lambda) \int dz \ p(z|x,\lambda) \tau_{z,\lambda} \qquad \forall x.
	\end{equation}
	If this is possible, we say that $\mathcal{E}$ is $r$-simulable. Notably, we can define $\sigma_{x,\lambda}\equiv \int dz p(z|x,\lambda) \tau_{z,\lambda}$ and notice that each $\sigma_{x,\lambda}$ is a valid quantum state and that it is confined to the space $\Pi_\lambda$. Thus, we can without loss of generality ignore the postprocessing and consider that each $\mathcal{P}_\lambda$ emits an $m$-state ensemble in $\Pi_\lambda$. This leads us to a simpler definition of the quantum absolute dimension.  
	
	\begin{definition}[Quantum Absolute Dimension]\label{DefQAD}
		Let $\mathcal{E}=\{\rho_x\}_{x=1}^m$ be an ensemble of $d$-dimensional states. The absolute dimension of the ensemble is
		\begin{align}\label{QAD}
		r_Q(\mathcal{E})\equiv \min_{q,\sigma} \Big\{&r:  \quad \rho_x=\int d\lambda \, q(\lambda) \, \sigma_{x,\lambda}\\\nonumber
		&\text{where}\quad\exists  \Pi_\lambda  \hspace{3mm} \stt \quad \Pi_\lambda^2=\Pi_\lambda,\\\nonumber
		& \tr(\Pi_\lambda)=r, \quad \sigma_{x,\lambda}=\Pi_\lambda \sigma_{x,\lambda}\Pi_\lambda \quad \forall x  \Big\},
		\end{align} 
		where the minimisation is evaluated over the states $\{\sigma_{x,\lambda}\}$ and the probability density $\{q(\lambda)\}$, where $\lambda$ runs over all possible $r$-dimesional subspaces. 
	\end{definition}
	In general, the cardinality of the number of preparation devices, $\lambda$, can be arbitrary and possibly uncountable. We denote by $\mathcal{S}_{r,m}$ the set of  $r$-simulable ensembles of $m$ states. It is straightforward that $\mathcal{S}_{r,m}$ is convex and that $\mathcal{S}_{r,m}\subset \mathcal{S}_{r+1,m}$.
	
	To gain some basic understanding for the absolute dimension, we mention some simple special cases. If the absolute dimension is unit, i.e.~$r_Q=1$, the ensemble $\{\sigma_{x,\lambda}\}_x$ can only feature one unique state since $\Pi_\lambda=\ketbra{\phi_\lambda}$ is a pure state. Hence $\sigma_{x,\lambda}=\phi_\lambda$. Thus, the simulation \eqref{QAD} reduces to $\rho_x=\int d \lambda \, q(\lambda)\phi_\lambda$ which holds if and only if $\rho_1=\ldots=\rho_m$. Note that also a single mixed state has $r_Q=1$. Another special case is if $\E$ is comprised only of pure states, $\rho_x=\ketbra{\psi_x}$. Then $q(\lambda)$ must be deterministic and w.l.g.  the number of preparation devices  can be taken finite. This implies that $\ketbra{\psi_x}=\sum_{\lambda} D(\lambda)\,\sigma_{x,\lambda}$, where $\sigma_{x,\lambda}$ is pure and $D(\lambda)\in\{0,1\}$.  The  absolute dimension is then the number of linearly independent states in $\E$. 
	
	\textit{Witness methods.---} The fundamental question now is how to characterise the set $\mathcal{S}_{r,m}$. From a practical point of view, it is natural to consider witness-type criteria, that imply lower bounds on $r_Q$ by measuring a suitable observable. We now show how such witnesses can be systematically constructed. 
	
	Consider that a preparation device emits some ensemble $\mathcal{E}$ whose absolute dimension we want to detect. For this purpose, we can perform on the ensemble some set of measurements,  $\lbrace M_{b|y} \rbrace$, where $y$ denotes the measurement choice and $b$ denotes the outcome. Since $\mathcal{S}_{r,m}$ is convex, we need only to consider linear witnesses, which take the form 
	\begin{equation}
	\label{QAD_witness}
	W \equiv 	 \sum_{b,x,y} c_{bxy} \tr(\rho_x M_{b|y})\stackrel{r\text{-sim}}{\leq} \beta_r,
	\end{equation}
	for some real coefficients $c_{bxy}$ which w.l.g.  can be assumed non-negative. Here, $\beta_r$ is a bound respected by all  $r$-simulable ensembles. Thus, a violation of this inequality implies that $\E\notin \mathcal{S}_{r,m}$ and hence also the lower bound $r_Q(\mathcal{E})\geq r+1$.  Our first result shows how to compute such witnesses.

	\begin{result}[Witness method]\label{ResWitness}
		Given a set of measurements $\{M_{b|y}\}$ and non-negative coefficients $c_{bxy}$, any $r$-simulable ensemble $\E$ satisfies the inequality
		\begin{equation}
		\label{Witness_inequality}
		W(\mathcal{E})\leq \sum_{i=1}^r \lambda_i\bigg(\sum_x O_x\bigg),
		\end{equation}
		where $\{\lambda_i(X)\}_i$ denotes the eigenvalues of $X$ in non-increasing order and $O_x=\sum_{b,y} c_{bxy} M_{b|y}$. 
	\end{result}
	\begin{proof}
		To show this, we first substitute $\rho_x$ with the definition in Eq.~\eqref{QAD}. Because of the linearity of the witness and the convexity of $\mathcal{S}_{r,m}$, we can w.l.g.  restrict to considering deterministic distributions $q(\lambda)$, i.e.~only one value $\lambda=\lambda^*$ is relevant for maximising $W(\mathcal{E})$. Then it holds that  $W=\sum_x\tr(\sigma_{x,\lambda^*}\Pi_{\lambda^*}O_x\Pi_{\lambda^*})\leq \sum_{x}\lambda_\text{max}(\Pi_{\lambda^*}O_x\Pi_{\lambda^*})\leq \tr(\Pi_{\lambda^*}\sum_x O_x)\leq\sum_{i=1}^r \lambda_i(\sum_x O_x)$ which is the result \eqref{Witness_inequality}. In the first step, we used that  $\Pi_{\lambda^*}\sigma_{x,\lambda^*}\Pi_{\lambda^*}=\sigma_{x,\lambda^*}$, then that the optimal state is the projection onto the eigenvector associated with the largest eigenvalue of $\Pi_{\lambda^*}O_x\Pi_{\lambda^*}$, then that $\lambda_\text{max}(A)\leq \tr(A)$ for $A\succeq 0$, and lastly that the optimal $\Pi_{\lambda^*}$ is the projector onto the eigenspace of $\sum_x O_x$ associated with the $r$ largest eigenvalues.  Additionally, to also saturate  the penultimate inequality, we need that $\Pi_{\lambda^*}O_x\Pi_{\lambda^*}$ is a rank-one operator.  Under these conditions, the inequality \eqref{Witness_inequality} is also tight.
	\end{proof}
	Note that, in particular, the inequality is tight whenever $O_x$ is a rank-one operator. To apply Result~\ref{ResWitness} to a target ensemble, one needs only to appropriately select the measurements and the coefficients of the witness. We will soon see that already very simple choices can be useful in practice.

	While Result~\ref{ResWitness} applies to general witnesses, a particularly interesting case is found when considering minimal error state discrimination  for an ensemble of $m\geq d$ states. This specific witness, $W_\text{disc}$, corresponds to the task of performing a single measurement (i.e.~$M_{b|y}=M_b$) and outputting $b=x$. Hence, the defining coefficients \eqref{QAD_witness} of $W_\text{disc}$ are $c_{bxy}=c_{bx}=\delta_{b,x}/m$. Then, from Result~\ref{ResWitness}, we get $\sum_x O_x=\frac{1}{m}\sum_x M_x=\frac{1}{m}\openone$ and hence Eq.~\eqref{Witness_inequality} becomes
	\begin{equation}\label{statediscrimination}
	W_\text{disc}\stackrel{r\text{-sim}}{\leq} \frac{r}{m}.
	\end{equation}
	Importantly, this holds \textit{independently} of the choice of measurement.  With appropriate normalisation, the optimal state discrimination probability can be interpreted as a one-shot accessible information quantity \cite{Konig2009}. The quantity is $I(\E)=H_\text{min}(\frac{1}{m})-H_\text{min}(\E)$, where $H_\text{min}(\frac{1}{m})=\log_2(m)$ is the min-entropy of a uniform prior and $H_\text{min}(\E)=-\log_2(W_\text{disc})$ is the observer's uncertainty about the classical label when provided the ensemble $\mathcal{E}$. Thus, we conclude that the absolute dimension implies a limitation on the accessible information.
	
	\begin{observation}
		Any ensemble with absolute dimension $r_Q$ carries at most $I=\log_2(r_Q)$ bits of accessible information. 
	\end{observation}
	This observation has a notable practical advantage: whether an ensemble violates \eqref{statediscrimination}, which is one-to-one with a violation of the above accessible information limit, can be computed efficiently as a semidefinite program over the measurement $\{M_b\}$ (see \cite{cobucci_2024_13882182} for implementation). This provides a simple way of finding sufficient conditions for showing that $\mathcal{E}\notin \mathcal{S}_{r,m}$ without having to adress how to best select the measurements used in the witness construction used in Result~\ref{ResWitness}.

	Eq.~\eqref{statediscrimination} allows us to obtain a simple and general bound on the white noise tolerance of an arbitrary high-dimensional quantum ensemble comprised of pure states.  Consider that with probability $v$ we generate some ensemble of $d$-dimensional pure states, $\{\ket{\psi_x}\}$,  and with probability $1-v$ we generate maximally mixed states;  $\rho_x=v\ketbra{\psi_x}+\frac{1-v}{d}\openone$. It follows that $W_\text{disc}(\{\rho_x\})=\max_{M} \frac{1}{m}\sum_{x}\tr(\rho_xM_x)= vW_\text{disc}(\{\psi_x\})+\frac{1-v}{m}\leq \frac{v(d-1)+1}{m}$. In the last step we used that $W_\text{disc}(\{\psi_x\})\leq \frac{d}{m}$ for any $d$-dimensional ensemble \cite{proof}. The critical value of $v$ for violating Eq.~\eqref{statediscrimination} then becomes

	\begin{equation}
	\label{Critical_visibility}
	v_{\text{crit}} = \frac{r-1}{d-1}.
	\end{equation}
	Thus, any isotropic $d$-dimensional quantum ensemble reaching the maximum discrimination and with a visibility exceeding this rate must have an absolute dimension of at least $r+1$. Notice that this result holds independently of the specific states appearing in the ensemble and independently of its size, $m$.

	\textit{Simulation models.---} The above criteria give necessary conditions for $r$-simulability but they are, in general, not sufficient conditions. We now formulate relevant sufficient conditions by constructing explicit $r$-dimensional simulation models for any ensemble of pure states subject to isotropic noise. As we will see, these models are even optimal for large classes of quantum ensembles. The following result provides bounds on the critical visibility of an ensemble, below which it is necessarily $r$-simulable. 
	\begin{result}[Simulation under isotropic noise]\label{ResWernerModel}
		Consider any $m$-state ensemble, $\E$, comprised of $d$-dimensional pure states mixed with isotropic noise, i.e.~$\rho_x=v\ketbra{\psi_x}+\frac{1-v}{d}\openone$, for some visibility $v\in[0,1]$. An $r$-dimensional simulation is possible whenever 
		\begin{equation}
		\label{Critical_visibility_sim}
		v\leq \frac{r-1}{d-1}.
		\end{equation}
		Moreover, for an ensemble with  $m\leq d$ states, an $r$-dimensional simulation is possible whenever 
		\begin{equation}\label{Critical_visibility_mdim}
		v \leq \frac{r-1}{d-1}\times\left(1+\frac{d-m}{m}\frac{d(m-r)+r}{d(m-r)+r-1}\right).
		\end{equation}
	\end{result}
	
	\begin{proof}
		Here, we sketch the simulation model leading to Eq.~\eqref{Critical_visibility_sim} and refer to Appendix \ref{Appendix_simulation} for complete details. Following Definition~\ref{DefQAD}, we consider a specific choice of projectors $\Pi_\lambda$. Every basis can be obtained by applying a $d$-dimensional unitary $U$ to the computational basis $\lbrace \ketbra{i} \rbrace_{i=1}^{d}$. We then select the $r$ first elements of the basis and define the projector  $\Pi_U=\sum_{i=1}^r U\ketbra{i} U^\dagger \equiv \Pi_\lambda$. Each $\rho_x$ is then simulated by averaging over the Haar measure of the normalised projection $\Pi_U\ket{\psi_x}$. The resulting integral over the Haar measure is computed using the techniques presented in \cite{Wiseman_2007,Jones_2007}, which leads to Eq.~\eqref{Critical_visibility_sim}. If one uses that the ensemble has fewer than $d$ states, an analogous protocol obtains the improved bounds in Eq.~\eqref{Critical_visibility_mdim}.
		
	\end{proof}
	
	We make some remarks on Result~\ref{ResWernerModel}. Firstly, note that Eq.~\eqref{Critical_visibility_sim} holds independently of $m$. Secondly, this upper bound matches the lower bound in Eq.~\eqref{Critical_visibility}. Thus, if $\{\ket{\psi_x}\}_{x=1}^m$ can be discriminated optimally, i.e.~with $W_\text{disc}=\frac{d}{m}$, the result is both necessary and sufficient. Natural examples of such ensembles are those that correspond to the rank-one projectors in a measurement. Thirdly, notice that choosing $m=d$ in Eq.~\eqref{Critical_visibility_mdim} reduces it to Eq.~\eqref{Critical_visibility_sim}. This highlights the property that a maximal isotropic noise tolerance is obtained already with a $d$-state ensemble. Fourthly, our  model  uses uncountably many $r$-dimensional subspaces to simulate $\mathcal{E}$ but this is not always necessary. In Appendix \ref{Appendix_finite-sim}, we show that for any ensemble in which $\{\ket{\psi_x}\}_{x=1}^m$ are orthonormal, the same bounds on the visibility as given in Result~\ref{ResWernerModel} can be achieved using only $\binom{d}{r}$ different $r$-dimensional subspaces. Finally, using a protocol analogous to the one presented above, it is possible to address also the case of ensembles with arbitrary $m$ but where $\{\ket{\psi_x}\}$ is confined to an $s$-dimensional subspace, with $s\leq d$. In Appendix \ref{Appendix_s_smaller_d-case}, we show that a simulation then is possible for $v \leq \frac{r-1}{d-1-r(d/s-1)}$. If we set $s=m$ this bound is worse than the one in Eq.~\eqref{Critical_visibility_mdim}, but it applies to more general ensembles.

	\begin{table}[t!]
		\begin{tabular}{|c|c|c|c|c|}
			\hline
			$\,$ r $\,$ & $\,$ Numerical $\,$ & $\,$  $\,$ Analytical $\,$ $\,$ & $\,$  $\,$ Difference  $\,$ $\,$ \\
			\hline
			2 & $0.1537$ & $0.1429$ & $0.0109$ \\
			\hline
			3 & $0.3099$ & $0.2857$ &  $0.0242$ \\
			\hline
			4 & $0.4647$ & $0.4286$ & $0.0361$ \\
			\hline
			5 & $0.6133$ & $0.5714$ & $0.0419$ \\
			\hline
			6 & $0.7525$ & $0.7143$ & $0.0382$ \\
			\hline
			7 & $0.8800$ & $0.8571$ & $0.0229$ \\
			\hline
		\end{tabular}
		\caption{Critical visibility, $v$, for $r$-simulating the $8$-dimensional ensemble consisting of all but the last computational basis state and the uniform superposition state. Both the analytical result from Eq.~\eqref{Critical_visibility_sim} and the numerical result from evaluating the SDP \eqref{SDP} are shown. The latter is evaluated using only $r$-dimensional subspaces selected from the computational and Fourier basis. This simple choice is sufficient to outperform the analytical model.}
		\label{table:visibility-sim-sdp}
	\end{table}

	\textit{Numerical method.---} The above simulation models are valid for isotropic noise, and they are also not always optimal for such ensembles. We therefore develop general and  efficient numerical tools for constructing $r$-dimensional simulations. Consider that we wish to simulate $\mathcal{E}$ using $r$-dimensional systems. To this end, let $\mathbf{b}_U=\{U\ket{i}\}_{i=1}^d$ be some basis of  $\mathbb{C}^d$, defined by the unitary $U$. In order to define our subspaces, we construct for each $\mathbf{b}_U$ the set $\mathbf{t}_U$, corresponding to all selections of $r$ vectors from the basis. Thus, the elements of $\mathbf{t}_U$ can be numbered $\mu=1,\ldots,\binom{d}{r}$ and to each we associate the rank-$r$ projector $ \Pi_{(U,\mu)}$. Thus, we aim to construct a model for $\mathcal{E}$ by restricting ourselves to the subspaces $\{\Pi_{(U,\mu)}\}_{U,\mu}$, where $U$ runs over a fixed finite selection of unitaries, $\mathcal{U}$. Whether a simulation is possible can be determined by the following semidefinite program
	\begin{equation}
	\label{SDP}
	\begin{aligned}
	\max_{v,\tilde{\sigma}} & \quad v\\
	\text{s.t.}&\quad v\rho_x+\frac{1-v}{d}\openone =  \sum_{U\in\mathcal{U}}\sum_{\mu=1}^{\binom{d}{r}} \tilde{\sigma}_{x,(U,\mu)}, \quad \forall \, x\\
	& \quad \tilde{\sigma}_{x,(U,\mu)} \geq 0 \quad \forall x,U,\mu,\\
	& \quad \tr(\tilde{\sigma}_{x,(U,\mu)}) = q(U,\mu), \quad q(U,\mu) \geq 0, \quad \forall x,U,\mu,\\
	& \quad \sum_{U \in \mathcal{U}} \sum_{\mu = 1}^{d \choose r} q(U,\mu) = 1, \\
	& \quad \tilde{\sigma}_{x,(U,\mu)}=\Pi_{(U,\mu)} \tilde{\sigma}_{x,(U,\mu)}\Pi_{(U,\mu)} \quad \forall x,U,\mu.
	\end{aligned}
	\end{equation}
	Here, the tuple $(U,\mu)$ plays the role of $\lambda$ in Definition~\ref{DefQAD} and $\tilde{\sigma}_{x,(U,\mu)} = q(U,\mu)\,\sigma_{x,(U,\mu)}$ are unnormalised states. Also, we have introduced auxiliary isotropic noise on $\mathcal{E}$ as a convenient quantifier of the simulability. Note that this choice of quantifier does not restrict the generality of the method; a simulation of $\mathcal{E}$ is possible with the selected subspaces only if the solution is $v=1$.

	It turns out that simple choices of $\mathcal{U}$ can lead to well-performing simulation models for ensembles that are not optimally simulated by the analytical approach of Result~\ref{ResWernerModel}. For example, consider all but one of the computational basis states and the uniform superposition state, i.e.~$\ket{\psi_x}=\ket{x}$ for $x=1,\ldots,d-1$ and $\ket{\psi_d}=\frac{1}{\sqrt{d}}\sum_{x=1}^d\ket{x}$. We consider the isotropic mixture of these states, i.e.~$\mathcal{E}=\{\rho_x\}$ with $\rho_x=v\ketbra{\psi_x}+\frac{1-v}{d}\openone$, and our goal is to $r$-simulate the ensemble. As a simple example, we select  $\mathcal{U}=\{\openone,F\}$, where $F$ is the Fourier matrix. In the simplest non-trivial case, we choose $d=3$ and $r=2$;  computing \eqref{SDP} gives $v\approx 0.5909$, which otperforms the threshold $v=\frac{1}{2}$ obtained from the model in Eq.~\eqref{Critical_visibility_sim}. To showcase the efficiency of this method, we evaluate \eqref{SDP} with the same $\mathcal{U}$ for significantly higher dimensions; we choose $d=8$ and consider all values of $r$. The results are given in Tab~\ref{table:visibility-sim-sdp}, where they also are compared with the analytical value from Result~\ref{ResWernerModel}. For every $r$, we observe an improvement. Lastly, numerical exploration shows that by extending the set $\mathcal{U}$, even better simulation models are typically within reach. Our implementation is available at \cite{cobucci_2024_13880206}.

	\textit{Discussion.---} We have proposed a way of assigning a dimensionality to quantum ensembles in a basis-independent way and shown how this absolute dimension can be characterised. Some of our methods are conceptually motivated, e.g.~the simulation models, whereas others are practically well-suited for experimental tests, e.g.~the witness method. Our contribution can be placed in the context of works that take a perspective-independent approach to quantum resources, e.g.~entanglement \cite{Cai2021} and coherence \cite{Yao2015, Radhakrishnan2019, Designolle_2021}. 
	
	The possibility to address dimesionality in an absolute way provides a useful perspective on the advantages of high dimensional systems for quantum information processing. For instance, in quantum key distribution, high-dimensional ensembles are known to increase security and bit rates \cite{Cerf2002, Sheridan2010}. However, if the ensemble arriving to the receiver is sufficiently noisy and/or the channel sufficiently lossy, the $d$-dimensional ensemble will have a smaller absolute dimension. Thus, the eavesdropper could prepare the $r$-simulation of the ensemble for the receiver, and thus the protocol will be limited by the ultimate performance of $r$-dimensional protocols. From a practical point of view, there is often a trade-off between dimensionality and experimental control; see e.g.~\cite{Ecker2019, Reimer2019, Srivastav2022}, and the possibility to analyse the absolute dimension of noisy ensembles provides a tool to assess this trade-off and to better access the quantum information advantages of an increased dimesionality.

	\begin{acknowledgements}
		A.B and M.J.R  acknowledge the hospitality of the Lund quantum information group. We thank Roope Uola for discussions. This work is supported by the Wenner-Gren Foundation, by the Knut and Alice Wallenberg Foundation through the Wallenberg Center for Quantum Technology (WACQT)  and by the Swedish Research Council under Contract No. 2023-03498.  A.B.  acknowledges the support of the Spanish Agencia Estatal de Investigacion through the grants ``IFT Centro de Excelencia Severo Ochoa CEX2020-001007-S'', PID2019-110058GB-C22 and PID2022-142545NB-C22 funded by MCIN/AEI/10.13039/501100011033 and by ERDF. The work of A.B. is supported through the FPI grant PRE2020-095867 funded by MCIN/AEI/10.13039/501100011033. M.J.R. acknowledges financial support by the Vienna Doctoral School in Physics (VDSP). This research was funded in whole, or in part, by the Austrian Science Fund (FWF) through BeyondC with Grant-DOI 10.55776/F71.
	\end{acknowledgements}

	\twocolumngrid
	\bibliography{bibliography.bib}

\begin{thebibliography}{32}%
\makeatletter
\providecommand \@ifxundefined [1]{%
 \@ifx{#1\undefined}
}%
\providecommand \@ifnum [1]{%
 \ifnum #1\expandafter \@firstoftwo
 \else \expandafter \@secondoftwo
 \fi
}%
\providecommand \@ifx [1]{%
 \ifx #1\expandafter \@firstoftwo
 \else \expandafter \@secondoftwo
 \fi
}%
\providecommand \natexlab [1]{#1}%
\providecommand \enquote  [1]{``#1''}%
\providecommand \bibnamefont  [1]{#1}%
\providecommand \bibfnamefont [1]{#1}%
\providecommand \citenamefont [1]{#1}%
\providecommand \href@noop [0]{\@secondoftwo}%
\providecommand \href [0]{\begingroup \@sanitize@url \@href}%
\providecommand \@href[1]{\@@startlink{#1}\@@href}%
\providecommand \@@href[1]{\endgroup#1\@@endlink}%
\providecommand \@sanitize@url [0]{\catcode `\\12\catcode `\$12\catcode
  `\&12\catcode `\#12\catcode `\^12\catcode `\_12\catcode `\%12\relax}%
\providecommand \@@startlink[1]{}%
\providecommand \@@endlink[0]{}%
\providecommand \url  [0]{\begingroup\@sanitize@url \@url }%
\providecommand \@url [1]{\endgroup\@href {#1}{\urlprefix }}%
\providecommand \urlprefix  [0]{URL }%
\providecommand \Eprint [0]{\href }%
\providecommand \doibase [0]{https://doi.org/}%
\providecommand \selectlanguage [0]{\@gobble}%
\providecommand \bibinfo  [0]{\@secondoftwo}%
\providecommand \bibfield  [0]{\@secondoftwo}%
\providecommand \translation [1]{[#1]}%
\providecommand \BibitemOpen [0]{}%
\providecommand \bibitemStop [0]{}%
\providecommand \bibitemNoStop [0]{.\EOS\space}%
\providecommand \EOS [0]{\spacefactor3000\relax}%
\providecommand \BibitemShut  [1]{\csname bibitem#1\endcsname}%
\let\auto@bib@innerbib\@empty
\bibitem [{\citenamefont {Terhal}\ and\ \citenamefont
  {Horodecki}(2000)}]{Terhal2000}%
  \BibitemOpen
  \bibfield  {author} {\bibinfo {author} {\bibfnamefont {B.~M.}\ \bibnamefont
  {Terhal}}\ and\ \bibinfo {author} {\bibfnamefont {P.}~\bibnamefont
  {Horodecki}},\ }\bibfield  {title} {\bibinfo {title} {Schmidt number for
  density matrices},\ }\href {https://doi.org/10.1103/PhysRevA.61.040301}
  {\bibfield  {journal} {\bibinfo  {journal} {Phys. Rev. A}\ }\textbf {\bibinfo
  {volume} {61}},\ \bibinfo {pages} {040301} (\bibinfo {year}
  {2000})}\BibitemShut {NoStop}%
\bibitem [{\citenamefont {Wolf}\ and\ \citenamefont
  {Perez-Garcia}(2009)}]{Wolf2009}%
  \BibitemOpen
  \bibfield  {author} {\bibinfo {author} {\bibfnamefont {M.~M.}\ \bibnamefont
  {Wolf}}\ and\ \bibinfo {author} {\bibfnamefont {D.}~\bibnamefont
  {Perez-Garcia}},\ }\bibfield  {title} {\bibinfo {title} {Assessing quantum
  dimensionality from observable dynamics},\ }\href
  {https://doi.org/10.1103/PhysRevLett.102.190504} {\bibfield  {journal}
  {\bibinfo  {journal} {Phys. Rev. Lett.}\ }\textbf {\bibinfo {volume} {102}},\
  \bibinfo {pages} {190504} (\bibinfo {year} {2009})}\BibitemShut {NoStop}%
\bibitem [{\citenamefont {Ioannou}\ \emph {et~al.}(2022)\citenamefont
  {Ioannou}, \citenamefont {Sekatski}, \citenamefont {Designolle},
  \citenamefont {Jones}, \citenamefont {Uola},\ and\ \citenamefont
  {Brunner}}]{Ioannou2022}%
  \BibitemOpen
  \bibfield  {author} {\bibinfo {author} {\bibfnamefont {M.}~\bibnamefont
  {Ioannou}}, \bibinfo {author} {\bibfnamefont {P.}~\bibnamefont {Sekatski}},
  \bibinfo {author} {\bibfnamefont {S.}~\bibnamefont {Designolle}}, \bibinfo
  {author} {\bibfnamefont {B.~D.~M.}\ \bibnamefont {Jones}}, \bibinfo {author}
  {\bibfnamefont {R.}~\bibnamefont {Uola}},\ and\ \bibinfo {author}
  {\bibfnamefont {N.}~\bibnamefont {Brunner}},\ }\bibfield  {title} {\bibinfo
  {title} {Simulability of high-dimensional quantum measurements},\ }\href
  {https://doi.org/10.1103/PhysRevLett.129.190401} {\bibfield  {journal}
  {\bibinfo  {journal} {Phys. Rev. Lett.}\ }\textbf {\bibinfo {volume} {129}},\
  \bibinfo {pages} {190401} (\bibinfo {year} {2022})}\BibitemShut {NoStop}%
\bibitem [{\citenamefont {Navascu\'es}\ \emph {et~al.}(2014)\citenamefont
  {Navascu\'es}, \citenamefont {de~la Torre},\ and\ \citenamefont
  {V\'ertesi}}]{Navascues2014}%
  \BibitemOpen
  \bibfield  {author} {\bibinfo {author} {\bibfnamefont {M.}~\bibnamefont
  {Navascu\'es}}, \bibinfo {author} {\bibfnamefont {G.}~\bibnamefont {de~la
  Torre}},\ and\ \bibinfo {author} {\bibfnamefont {T.}~\bibnamefont
  {V\'ertesi}},\ }\bibfield  {title} {\bibinfo {title} {Characterization of
  quantum correlations with local dimension constraints and its
  device-independent applications},\ }\href
  {https://doi.org/10.1103/PhysRevX.4.011011} {\bibfield  {journal} {\bibinfo
  {journal} {Phys. Rev. X}\ }\textbf {\bibinfo {volume} {4}},\ \bibinfo {pages}
  {011011} (\bibinfo {year} {2014})}\BibitemShut {NoStop}%
\bibitem [{\citenamefont {Cozzolino}\ \emph {et~al.}(2019)\citenamefont
  {Cozzolino}, \citenamefont {Da~Lio}, \citenamefont {Bacco},\ and\
  \citenamefont {Oxenløwe}}]{Cozzolino2019}%
  \BibitemOpen
  \bibfield  {author} {\bibinfo {author} {\bibfnamefont {D.}~\bibnamefont
  {Cozzolino}}, \bibinfo {author} {\bibfnamefont {B.}~\bibnamefont {Da~Lio}},
  \bibinfo {author} {\bibfnamefont {D.}~\bibnamefont {Bacco}},\ and\ \bibinfo
  {author} {\bibfnamefont {L.~K.}\ \bibnamefont {Oxenløwe}},\ }\bibfield
  {title} {\bibinfo {title} {High-dimensional quantum communication: Benefits,
  progress, and future challenges},\ }\href
  {https://doi.org/https://doi.org/10.1002/qute.201900038} {\bibfield
  {journal} {\bibinfo  {journal} {Advanced Quantum Technologies}\ }\textbf
  {\bibinfo {volume} {2}},\ \bibinfo {pages} {1900038} (\bibinfo {year}
  {2019})},\ \Eprint
  {https://arxiv.org/abs/https://onlinelibrary.wiley.com/doi/pdf/10.1002/qute.201900038}
  {https://onlinelibrary.wiley.com/doi/pdf/10.1002/qute.201900038} \BibitemShut
  {NoStop}%
\bibitem [{\citenamefont {Erhard}\ \emph {et~al.}(2020)\citenamefont {Erhard},
  \citenamefont {Krenn},\ and\ \citenamefont {Zeilinger}}]{Erhard2020}%
  \BibitemOpen
  \bibfield  {author} {\bibinfo {author} {\bibfnamefont {M.}~\bibnamefont
  {Erhard}}, \bibinfo {author} {\bibfnamefont {M.}~\bibnamefont {Krenn}},\ and\
  \bibinfo {author} {\bibfnamefont {A.}~\bibnamefont {Zeilinger}},\ }\bibfield
  {title} {\bibinfo {title} {Advances in high-dimensional quantum
  entanglement},\ }\href {https://doi.org/10.1038/s42254-020-0193-5} {\bibfield
   {journal} {\bibinfo  {journal} {Nature Reviews Physics}\ }\textbf {\bibinfo
  {volume} {2}},\ \bibinfo {pages} {365} (\bibinfo {year} {2020})}\BibitemShut
  {NoStop}%
\bibitem [{\citenamefont {Wang}\ \emph {et~al.}(2020)\citenamefont {Wang},
  \citenamefont {Hu}, \citenamefont {Sanders},\ and\ \citenamefont
  {Kais}}]{Wang2020}%
  \BibitemOpen
  \bibfield  {author} {\bibinfo {author} {\bibfnamefont {Y.}~\bibnamefont
  {Wang}}, \bibinfo {author} {\bibfnamefont {Z.}~\bibnamefont {Hu}}, \bibinfo
  {author} {\bibfnamefont {B.~C.}\ \bibnamefont {Sanders}},\ and\ \bibinfo
  {author} {\bibfnamefont {S.}~\bibnamefont {Kais}},\ }\bibfield  {title}
  {\bibinfo {title} {Qudits and high-dimensional quantum computing},\
  }\bibfield  {journal} {\bibinfo  {journal} {Frontiers in Physics}\ }\textbf
  {\bibinfo {volume} {8}},\ \href {https://doi.org/10.3389/fphy.2020.589504}
  {10.3389/fphy.2020.589504} (\bibinfo {year} {2020})\BibitemShut {NoStop}%
\bibitem [{\citenamefont {Tavakoli}\ \emph {et~al.}(2018)\citenamefont
  {Tavakoli}, \citenamefont {Kaniewski}, \citenamefont {V\'ertesi},
  \citenamefont {Rosset},\ and\ \citenamefont {Brunner}}]{Tavakoli2018}%
  \BibitemOpen
  \bibfield  {author} {\bibinfo {author} {\bibfnamefont {A.}~\bibnamefont
  {Tavakoli}}, \bibinfo {author} {\bibfnamefont {J.~m.~k.}\ \bibnamefont
  {Kaniewski}}, \bibinfo {author} {\bibfnamefont {T.}~\bibnamefont
  {V\'ertesi}}, \bibinfo {author} {\bibfnamefont {D.}~\bibnamefont {Rosset}},\
  and\ \bibinfo {author} {\bibfnamefont {N.}~\bibnamefont {Brunner}},\
  }\bibfield  {title} {\bibinfo {title} {Self-testing quantum states and
  measurements in the prepare-and-measure scenario},\ }\href
  {https://doi.org/10.1103/PhysRevA.98.062307} {\bibfield  {journal} {\bibinfo
  {journal} {Phys. Rev. A}\ }\textbf {\bibinfo {volume} {98}},\ \bibinfo
  {pages} {062307} (\bibinfo {year} {2018})}\BibitemShut {NoStop}%
\bibitem [{\citenamefont {Navascu\'es}\ \emph {et~al.}(2023)\citenamefont
  {Navascu\'es}, \citenamefont {P\'al}, \citenamefont {V\'ertesi},\ and\
  \citenamefont {Ara\'ujo}}]{Navascues2023}%
  \BibitemOpen
  \bibfield  {author} {\bibinfo {author} {\bibfnamefont {M.}~\bibnamefont
  {Navascu\'es}}, \bibinfo {author} {\bibfnamefont {K.~F.}\ \bibnamefont
  {P\'al}}, \bibinfo {author} {\bibfnamefont {T.}~\bibnamefont {V\'ertesi}},\
  and\ \bibinfo {author} {\bibfnamefont {M.}~\bibnamefont {Ara\'ujo}},\
  }\bibfield  {title} {\bibinfo {title} {Self-testing in prepare-and-measure
  scenarios and a robust version of wigner's theorem},\ }\href
  {https://doi.org/10.1103/PhysRevLett.131.250802} {\bibfield  {journal}
  {\bibinfo  {journal} {Phys. Rev. Lett.}\ }\textbf {\bibinfo {volume} {131}},\
  \bibinfo {pages} {250802} (\bibinfo {year} {2023})}\BibitemShut {NoStop}%
\bibitem [{\citenamefont {Farkas}\ and\ \citenamefont
  {Kaniewski}(2019)}]{Farkas2019}%
  \BibitemOpen
  \bibfield  {author} {\bibinfo {author} {\bibfnamefont {M.}~\bibnamefont
  {Farkas}}\ and\ \bibinfo {author} {\bibfnamefont {J.~m.~k.}\ \bibnamefont
  {Kaniewski}},\ }\bibfield  {title} {\bibinfo {title} {Self-testing mutually
  unbiased bases in the prepare-and-measure scenario},\ }\href
  {https://doi.org/10.1103/PhysRevA.99.032316} {\bibfield  {journal} {\bibinfo
  {journal} {Phys. Rev. A}\ }\textbf {\bibinfo {volume} {99}},\ \bibinfo
  {pages} {032316} (\bibinfo {year} {2019})}\BibitemShut {NoStop}%
\bibitem [{\citenamefont {Tavakoli}\ \emph {et~al.}(2019)\citenamefont
  {Tavakoli}, \citenamefont {Rosset},\ and\ \citenamefont
  {Renou}}]{Rosset2019}%
  \BibitemOpen
  \bibfield  {author} {\bibinfo {author} {\bibfnamefont {A.}~\bibnamefont
  {Tavakoli}}, \bibinfo {author} {\bibfnamefont {D.}~\bibnamefont {Rosset}},\
  and\ \bibinfo {author} {\bibfnamefont {M.-O.}\ \bibnamefont {Renou}},\
  }\bibfield  {title} {\bibinfo {title} {Enabling computation of correlation
  bounds for finite-dimensional quantum systems via symmetrization},\ }\href
  {https://doi.org/10.1103/PhysRevLett.122.070501} {\bibfield  {journal}
  {\bibinfo  {journal} {Phys. Rev. Lett.}\ }\textbf {\bibinfo {volume} {122}},\
  \bibinfo {pages} {070501} (\bibinfo {year} {2019})}\BibitemShut {NoStop}%
\bibitem [{\citenamefont {Gallego}\ \emph {et~al.}(2010)\citenamefont
  {Gallego}, \citenamefont {Brunner}, \citenamefont {Hadley},\ and\
  \citenamefont {Ac\'{\i}n}}]{Gallego2010}%
  \BibitemOpen
  \bibfield  {author} {\bibinfo {author} {\bibfnamefont {R.}~\bibnamefont
  {Gallego}}, \bibinfo {author} {\bibfnamefont {N.}~\bibnamefont {Brunner}},
  \bibinfo {author} {\bibfnamefont {C.}~\bibnamefont {Hadley}},\ and\ \bibinfo
  {author} {\bibfnamefont {A.}~\bibnamefont {Ac\'{\i}n}},\ }\bibfield  {title}
  {\bibinfo {title} {Device-independent tests of classical and quantum
  dimensions},\ }\href {https://doi.org/10.1103/PhysRevLett.105.230501}
  {\bibfield  {journal} {\bibinfo  {journal} {Phys. Rev. Lett.}\ }\textbf
  {\bibinfo {volume} {105}},\ \bibinfo {pages} {230501} (\bibinfo {year}
  {2010})}\BibitemShut {NoStop}%
\bibitem [{\citenamefont {Ahrens}\ \emph {et~al.}(2012)\citenamefont {Ahrens},
  \citenamefont {Badziag}, \citenamefont {Cabello},\ and\ \citenamefont
  {Bourennane}}]{Ahrens2012}%
  \BibitemOpen
  \bibfield  {author} {\bibinfo {author} {\bibfnamefont {J.}~\bibnamefont
  {Ahrens}}, \bibinfo {author} {\bibfnamefont {P.}~\bibnamefont {Badziag}},
  \bibinfo {author} {\bibfnamefont {A.}~\bibnamefont {Cabello}},\ and\ \bibinfo
  {author} {\bibfnamefont {M.}~\bibnamefont {Bourennane}},\ }\bibfield  {title}
  {\bibinfo {title} {Experimental device-independent tests of classical and
  quantum dimensions},\ }\href {https://doi.org/10.1038/nphys2333} {\bibfield
  {journal} {\bibinfo  {journal} {Nature Physics}\ }\textbf {\bibinfo {volume}
  {8}},\ \bibinfo {pages} {592} (\bibinfo {year} {2012})}\BibitemShut {NoStop}%
\bibitem [{\citenamefont {Hendrych}\ \emph {et~al.}(2012)\citenamefont
  {Hendrych}, \citenamefont {Gallego}, \citenamefont {Mi{\v{c}}uda},
  \citenamefont {Brunner}, \citenamefont {Ac{\'i}n},\ and\ \citenamefont
  {Torres}}]{Hendrych2012}%
  \BibitemOpen
  \bibfield  {author} {\bibinfo {author} {\bibfnamefont {M.}~\bibnamefont
  {Hendrych}}, \bibinfo {author} {\bibfnamefont {R.}~\bibnamefont {Gallego}},
  \bibinfo {author} {\bibfnamefont {M.}~\bibnamefont {Mi{\v{c}}uda}}, \bibinfo
  {author} {\bibfnamefont {N.}~\bibnamefont {Brunner}}, \bibinfo {author}
  {\bibfnamefont {A.}~\bibnamefont {Ac{\'i}n}},\ and\ \bibinfo {author}
  {\bibfnamefont {J.~P.}\ \bibnamefont {Torres}},\ }\bibfield  {title}
  {\bibinfo {title} {Experimental estimation of the dimension of classical and
  quantum systems},\ }\href {https://doi.org/10.1038/nphys2334} {\bibfield
  {journal} {\bibinfo  {journal} {Nature Physics}\ }\textbf {\bibinfo {volume}
  {8}},\ \bibinfo {pages} {588} (\bibinfo {year} {2012})}\BibitemShut {NoStop}%
\bibitem [{\citenamefont {Tavakoli}\ \emph {et~al.}(2020)\citenamefont
  {Tavakoli}, \citenamefont {Zambrini~Cruzeiro}, \citenamefont {Bohr~Brask},
  \citenamefont {Gisin},\ and\ \citenamefont
  {Brunner}}]{Tavakoli2020informationally}%
  \BibitemOpen
  \bibfield  {author} {\bibinfo {author} {\bibfnamefont {A.}~\bibnamefont
  {Tavakoli}}, \bibinfo {author} {\bibfnamefont {E.}~\bibnamefont
  {Zambrini~Cruzeiro}}, \bibinfo {author} {\bibfnamefont {J.}~\bibnamefont
  {Bohr~Brask}}, \bibinfo {author} {\bibfnamefont {N.}~\bibnamefont {Gisin}},\
  and\ \bibinfo {author} {\bibfnamefont {N.}~\bibnamefont {Brunner}},\
  }\bibfield  {title} {\bibinfo {title} {Informationally restricted quantum
  correlations},\ }\href {https://doi.org/10.22331/q-2020-09-24-332} {\bibfield
   {journal} {\bibinfo  {journal} {{Quantum}}\ }\textbf {\bibinfo {volume}
  {4}},\ \bibinfo {pages} {332} (\bibinfo {year} {2020})}\BibitemShut {NoStop}%
\bibitem [{\citenamefont {Tavakoli}\ \emph {et~al.}(2022)\citenamefont
  {Tavakoli}, \citenamefont {Zambrini~Cruzeiro}, \citenamefont {Woodhead},\
  and\ \citenamefont {Pironio}}]{Tavakoli2022informationally}%
  \BibitemOpen
  \bibfield  {author} {\bibinfo {author} {\bibfnamefont {A.}~\bibnamefont
  {Tavakoli}}, \bibinfo {author} {\bibfnamefont {E.}~\bibnamefont
  {Zambrini~Cruzeiro}}, \bibinfo {author} {\bibfnamefont {E.}~\bibnamefont
  {Woodhead}},\ and\ \bibinfo {author} {\bibfnamefont {S.}~\bibnamefont
  {Pironio}},\ }\bibfield  {title} {\bibinfo {title} {Informationally
  restricted correlations: a general framework for classical and quantum
  systems},\ }\href {https://doi.org/10.22331/q-2022-01-05-620} {\bibfield
  {journal} {\bibinfo  {journal} {{Quantum}}\ }\textbf {\bibinfo {volume}
  {6}},\ \bibinfo {pages} {620} (\bibinfo {year} {2022})}\BibitemShut {NoStop}%
\bibitem [{\citenamefont {Tavakoli}\ \emph {et~al.}(2024)\citenamefont
  {Tavakoli}, \citenamefont {Pozas-Kerstjens}, \citenamefont {Brown},\ and\
  \citenamefont
  {Araújo}}]{tavakoli2024semidefiniteprogrammingrelaxationsquantum}%
  \BibitemOpen
  \bibfield  {author} {\bibinfo {author} {\bibfnamefont {A.}~\bibnamefont
  {Tavakoli}}, \bibinfo {author} {\bibfnamefont {A.}~\bibnamefont
  {Pozas-Kerstjens}}, \bibinfo {author} {\bibfnamefont {P.}~\bibnamefont
  {Brown}},\ and\ \bibinfo {author} {\bibfnamefont {M.}~\bibnamefont
  {Araújo}},\ }\href {https://arxiv.org/abs/2307.02551} {\bibinfo {title}
  {Semidefinite programming relaxations for quantum correlations}} (\bibinfo
  {year} {2024}),\ \Eprint {https://arxiv.org/abs/2307.02551} {arXiv:2307.02551
  [quant-ph]} \BibitemShut {NoStop}%
\bibitem [{\citenamefont {Konig}\ \emph {et~al.}(2009)\citenamefont {Konig},
  \citenamefont {Renner},\ and\ \citenamefont {Schaffner}}]{Konig2009}%
  \BibitemOpen
  \bibfield  {author} {\bibinfo {author} {\bibfnamefont {R.}~\bibnamefont
  {Konig}}, \bibinfo {author} {\bibfnamefont {R.}~\bibnamefont {Renner}},\ and\
  \bibinfo {author} {\bibfnamefont {C.}~\bibnamefont {Schaffner}},\ }\bibfield
  {title} {\bibinfo {title} {The operational meaning of min- and max-entropy},\
  }\href {https://doi.org/10.1109/TIT.2009.2025545} {\bibfield  {journal}
  {\bibinfo  {journal} {IEEE Transactions on Information Theory}\ }\textbf
  {\bibinfo {volume} {55}},\ \bibinfo {pages} {4337} (\bibinfo {year}
  {2009})}\BibitemShut {NoStop}%
\bibitem [{\citenamefont {Cobucci}\ and\ \citenamefont
  {Tavakoli}(2024{\natexlab{a}})}]{cobucci_2024_13882182}%
  \BibitemOpen
  \bibfield  {author} {\bibinfo {author} {\bibfnamefont {G.}~\bibnamefont
  {Cobucci}}\ and\ \bibinfo {author} {\bibfnamefont {A.}~\bibnamefont
  {Tavakoli}},\ }\href {https://doi.org/10.5281/zenodo.13882182} {\bibinfo
  {title} {{SDP for state discrimination and accessible information}}}
  (\bibinfo {year} {2024}{\natexlab{a}})\BibitemShut {NoStop}%
\bibitem [{pro()}]{proof}%
  \BibitemOpen
  \href@noop {} {\bibinfo {title} {{To show this, simply note that
  $W_\text{disc}(\{\psi_x\})=\max_M \frac{1}{m}\sum_x \tr(\psi_xM_x)\leq \max_M
  \frac{1}{m}\sum_x \lambda_\text{max}(M_x)\leq \max_M \frac{1}{m}\sum_x
  \tr(M_x)=\frac{d}{m}$.}}}\BibitemShut {Stop}%
\bibitem [{\citenamefont {Wiseman}\ \emph {et~al.}(2007)\citenamefont
  {Wiseman}, \citenamefont {Jones},\ and\ \citenamefont
  {Doherty}}]{Wiseman_2007}%
  \BibitemOpen
  \bibfield  {author} {\bibinfo {author} {\bibfnamefont {H.~M.}\ \bibnamefont
  {Wiseman}}, \bibinfo {author} {\bibfnamefont {S.~J.}\ \bibnamefont {Jones}},\
  and\ \bibinfo {author} {\bibfnamefont {A.~C.}\ \bibnamefont {Doherty}},\
  }\bibfield  {title} {\bibinfo {title} {Steering, entanglement, nonlocality,
  and the einstein-podolsky-rosen paradox},\ }\bibfield  {journal} {\bibinfo
  {journal} {Physical Review Letters}\ }\textbf {\bibinfo {volume} {98}},\
  \href {https://doi.org/10.1103/physrevlett.98.140402}
  {10.1103/physrevlett.98.140402} (\bibinfo {year} {2007})\BibitemShut
  {NoStop}%
\bibitem [{\citenamefont {Jones}\ \emph {et~al.}(2007)\citenamefont {Jones},
  \citenamefont {Wiseman},\ and\ \citenamefont {Doherty}}]{Jones_2007}%
  \BibitemOpen
  \bibfield  {author} {\bibinfo {author} {\bibfnamefont {S.~J.}\ \bibnamefont
  {Jones}}, \bibinfo {author} {\bibfnamefont {H.~M.}\ \bibnamefont {Wiseman}},\
  and\ \bibinfo {author} {\bibfnamefont {A.~C.}\ \bibnamefont {Doherty}},\
  }\bibfield  {title} {\bibinfo {title} {Entanglement, einstein-podolsky-rosen
  correlations, bell nonlocality, and steering},\ }\bibfield  {journal}
  {\bibinfo  {journal} {Physical Review A}\ }\textbf {\bibinfo {volume} {76}},\
  \href {https://doi.org/10.1103/physreva.76.052116}
  {10.1103/physreva.76.052116} (\bibinfo {year} {2007})\BibitemShut {NoStop}%
\bibitem [{\citenamefont {Cobucci}\ and\ \citenamefont
  {Tavakoli}(2024{\natexlab{b}})}]{cobucci_2024_13880206}%
  \BibitemOpen
  \bibfield  {author} {\bibinfo {author} {\bibfnamefont {G.}~\bibnamefont
  {Cobucci}}\ and\ \bibinfo {author} {\bibfnamefont {A.}~\bibnamefont
  {Tavakoli}},\ }\href {https://doi.org/10.5281/zenodo.13880206} {\bibinfo
  {title} {Qadsimulation}} (\bibinfo {year} {2024}{\natexlab{b}})\BibitemShut
  {NoStop}%
\bibitem [{\citenamefont {Cai}\ \emph {et~al.}(2021)\citenamefont {Cai},
  \citenamefont {Yu}, \citenamefont {Jayachandran}, \citenamefont {Brunner},
  \citenamefont {Scarani},\ and\ \citenamefont {Bancal}}]{Cai2021}%
  \BibitemOpen
  \bibfield  {author} {\bibinfo {author} {\bibfnamefont {Y.}~\bibnamefont
  {Cai}}, \bibinfo {author} {\bibfnamefont {B.}~\bibnamefont {Yu}}, \bibinfo
  {author} {\bibfnamefont {P.}~\bibnamefont {Jayachandran}}, \bibinfo {author}
  {\bibfnamefont {N.}~\bibnamefont {Brunner}}, \bibinfo {author} {\bibfnamefont
  {V.}~\bibnamefont {Scarani}},\ and\ \bibinfo {author} {\bibfnamefont {J.-D.}\
  \bibnamefont {Bancal}},\ }\bibfield  {title} {\bibinfo {title} {Entanglement
  for any definition of two subsystems},\ }\href
  {https://doi.org/10.1103/PhysRevA.103.052432} {\bibfield  {journal} {\bibinfo
   {journal} {Phys. Rev. A}\ }\textbf {\bibinfo {volume} {103}},\ \bibinfo
  {pages} {052432} (\bibinfo {year} {2021})}\BibitemShut {NoStop}%
\bibitem [{\citenamefont {Yao}\ \emph {et~al.}(2015)\citenamefont {Yao},
  \citenamefont {Xiao}, \citenamefont {Ge},\ and\ \citenamefont
  {Sun}}]{Yao2015}%
  \BibitemOpen
  \bibfield  {author} {\bibinfo {author} {\bibfnamefont {Y.}~\bibnamefont
  {Yao}}, \bibinfo {author} {\bibfnamefont {X.}~\bibnamefont {Xiao}}, \bibinfo
  {author} {\bibfnamefont {L.}~\bibnamefont {Ge}},\ and\ \bibinfo {author}
  {\bibfnamefont {C.~P.}\ \bibnamefont {Sun}},\ }\bibfield  {title} {\bibinfo
  {title} {Quantum coherence in multipartite systems},\ }\href
  {https://doi.org/10.1103/PhysRevA.92.022112} {\bibfield  {journal} {\bibinfo
  {journal} {Phys. Rev. A}\ }\textbf {\bibinfo {volume} {92}},\ \bibinfo
  {pages} {022112} (\bibinfo {year} {2015})}\BibitemShut {NoStop}%
\bibitem [{\citenamefont {Radhakrishnan}\ \emph {et~al.}(2019)\citenamefont
  {Radhakrishnan}, \citenamefont {Ding}, \citenamefont {Shi}, \citenamefont
  {Du},\ and\ \citenamefont {Byrnes}}]{Radhakrishnan2019}%
  \BibitemOpen
  \bibfield  {author} {\bibinfo {author} {\bibfnamefont {C.}~\bibnamefont
  {Radhakrishnan}}, \bibinfo {author} {\bibfnamefont {Z.}~\bibnamefont {Ding}},
  \bibinfo {author} {\bibfnamefont {F.}~\bibnamefont {Shi}}, \bibinfo {author}
  {\bibfnamefont {J.}~\bibnamefont {Du}},\ and\ \bibinfo {author}
  {\bibfnamefont {T.}~\bibnamefont {Byrnes}},\ }\bibfield  {title} {\bibinfo
  {title} {Basis-independent quantum coherence and its distribution},\ }\href
  {https://doi.org/https://doi.org/10.1016/j.aop.2019.04.020} {\bibfield
  {journal} {\bibinfo  {journal} {Annals of Physics}\ }\textbf {\bibinfo
  {volume} {409}},\ \bibinfo {pages} {167906} (\bibinfo {year}
  {2019})}\BibitemShut {NoStop}%
\bibitem [{\citenamefont {Designolle}\ \emph {et~al.}(2021)\citenamefont
  {Designolle}, \citenamefont {Uola}, \citenamefont {Luoma},\ and\
  \citenamefont {Brunner}}]{Designolle_2021}%
  \BibitemOpen
  \bibfield  {author} {\bibinfo {author} {\bibfnamefont {S.}~\bibnamefont
  {Designolle}}, \bibinfo {author} {\bibfnamefont {R.}~\bibnamefont {Uola}},
  \bibinfo {author} {\bibfnamefont {K.}~\bibnamefont {Luoma}},\ and\ \bibinfo
  {author} {\bibfnamefont {N.}~\bibnamefont {Brunner}},\ }\bibfield  {title}
  {\bibinfo {title} {Set coherence: Basis-independent quantification of quantum
  coherence},\ }\bibfield  {journal} {\bibinfo  {journal} {Physical Review
  Letters}\ }\textbf {\bibinfo {volume} {126}},\ \href
  {https://doi.org/10.1103/physrevlett.126.220404}
  {10.1103/physrevlett.126.220404} (\bibinfo {year} {2021})\BibitemShut
  {NoStop}%
\bibitem [{\citenamefont {Cerf}\ \emph {et~al.}(2002)\citenamefont {Cerf},
  \citenamefont {Bourennane}, \citenamefont {Karlsson},\ and\ \citenamefont
  {Gisin}}]{Cerf2002}%
  \BibitemOpen
  \bibfield  {author} {\bibinfo {author} {\bibfnamefont {N.~J.}\ \bibnamefont
  {Cerf}}, \bibinfo {author} {\bibfnamefont {M.}~\bibnamefont {Bourennane}},
  \bibinfo {author} {\bibfnamefont {A.}~\bibnamefont {Karlsson}},\ and\
  \bibinfo {author} {\bibfnamefont {N.}~\bibnamefont {Gisin}},\ }\bibfield
  {title} {\bibinfo {title} {Security of quantum key distribution using
  $\mathit{d}$-level systems},\ }\href
  {https://doi.org/10.1103/PhysRevLett.88.127902} {\bibfield  {journal}
  {\bibinfo  {journal} {Phys. Rev. Lett.}\ }\textbf {\bibinfo {volume} {88}},\
  \bibinfo {pages} {127902} (\bibinfo {year} {2002})}\BibitemShut {NoStop}%
\bibitem [{\citenamefont {Sheridan}\ and\ \citenamefont
  {Scarani}(2010)}]{Sheridan2010}%
  \BibitemOpen
  \bibfield  {author} {\bibinfo {author} {\bibfnamefont {L.}~\bibnamefont
  {Sheridan}}\ and\ \bibinfo {author} {\bibfnamefont {V.}~\bibnamefont
  {Scarani}},\ }\bibfield  {title} {\bibinfo {title} {Security proof for
  quantum key distribution using qudit systems},\ }\href
  {https://doi.org/10.1103/PhysRevA.82.030301} {\bibfield  {journal} {\bibinfo
  {journal} {Phys. Rev. A}\ }\textbf {\bibinfo {volume} {82}},\ \bibinfo
  {pages} {030301} (\bibinfo {year} {2010})}\BibitemShut {NoStop}%
\bibitem [{\citenamefont {Ecker}\ \emph {et~al.}(2019)\citenamefont {Ecker},
  \citenamefont {Bouchard}, \citenamefont {Bulla}, \citenamefont {Brandt},
  \citenamefont {Kohout}, \citenamefont {Steinlechner}, \citenamefont
  {Fickler}, \citenamefont {Malik}, \citenamefont {Guryanova}, \citenamefont
  {Ursin},\ and\ \citenamefont {Huber}}]{Ecker2019}%
  \BibitemOpen
  \bibfield  {author} {\bibinfo {author} {\bibfnamefont {S.}~\bibnamefont
  {Ecker}}, \bibinfo {author} {\bibfnamefont {F.}~\bibnamefont {Bouchard}},
  \bibinfo {author} {\bibfnamefont {L.}~\bibnamefont {Bulla}}, \bibinfo
  {author} {\bibfnamefont {F.}~\bibnamefont {Brandt}}, \bibinfo {author}
  {\bibfnamefont {O.}~\bibnamefont {Kohout}}, \bibinfo {author} {\bibfnamefont
  {F.}~\bibnamefont {Steinlechner}}, \bibinfo {author} {\bibfnamefont
  {R.}~\bibnamefont {Fickler}}, \bibinfo {author} {\bibfnamefont
  {M.}~\bibnamefont {Malik}}, \bibinfo {author} {\bibfnamefont
  {Y.}~\bibnamefont {Guryanova}}, \bibinfo {author} {\bibfnamefont
  {R.}~\bibnamefont {Ursin}},\ and\ \bibinfo {author} {\bibfnamefont
  {M.}~\bibnamefont {Huber}},\ }\bibfield  {title} {\bibinfo {title}
  {Overcoming noise in entanglement distribution},\ }\href
  {https://doi.org/10.1103/PhysRevX.9.041042} {\bibfield  {journal} {\bibinfo
  {journal} {Phys. Rev. X}\ }\textbf {\bibinfo {volume} {9}},\ \bibinfo {pages}
  {041042} (\bibinfo {year} {2019})}\BibitemShut {NoStop}%
\bibitem [{\citenamefont {Reimer}\ \emph {et~al.}(2019)\citenamefont {Reimer},
  \citenamefont {Sciara}, \citenamefont {Roztocki}, \citenamefont {Islam},
  \citenamefont {Romero~Cort{\'e}s}, \citenamefont {Zhang}, \citenamefont
  {Fischer}, \citenamefont {Loranger}, \citenamefont {Kashyap}, \citenamefont
  {Cino}, \citenamefont {Chu}, \citenamefont {Little}, \citenamefont {Moss},
  \citenamefont {Caspani}, \citenamefont {Munro}, \citenamefont {Aza{\~{n}}a},
  \citenamefont {Kues},\ and\ \citenamefont {Morandotti}}]{Reimer2019}%
  \BibitemOpen
  \bibfield  {author} {\bibinfo {author} {\bibfnamefont {C.}~\bibnamefont
  {Reimer}}, \bibinfo {author} {\bibfnamefont {S.}~\bibnamefont {Sciara}},
  \bibinfo {author} {\bibfnamefont {P.}~\bibnamefont {Roztocki}}, \bibinfo
  {author} {\bibfnamefont {M.}~\bibnamefont {Islam}}, \bibinfo {author}
  {\bibfnamefont {L.}~\bibnamefont {Romero~Cort{\'e}s}}, \bibinfo {author}
  {\bibfnamefont {Y.}~\bibnamefont {Zhang}}, \bibinfo {author} {\bibfnamefont
  {B.}~\bibnamefont {Fischer}}, \bibinfo {author} {\bibfnamefont
  {S.}~\bibnamefont {Loranger}}, \bibinfo {author} {\bibfnamefont
  {R.}~\bibnamefont {Kashyap}}, \bibinfo {author} {\bibfnamefont
  {A.}~\bibnamefont {Cino}}, \bibinfo {author} {\bibfnamefont {S.~T.}\
  \bibnamefont {Chu}}, \bibinfo {author} {\bibfnamefont {B.~E.}\ \bibnamefont
  {Little}}, \bibinfo {author} {\bibfnamefont {D.~J.}\ \bibnamefont {Moss}},
  \bibinfo {author} {\bibfnamefont {L.}~\bibnamefont {Caspani}}, \bibinfo
  {author} {\bibfnamefont {W.~J.}\ \bibnamefont {Munro}}, \bibinfo {author}
  {\bibfnamefont {J.}~\bibnamefont {Aza{\~{n}}a}}, \bibinfo {author}
  {\bibfnamefont {M.}~\bibnamefont {Kues}},\ and\ \bibinfo {author}
  {\bibfnamefont {R.}~\bibnamefont {Morandotti}},\ }\bibfield  {title}
  {\bibinfo {title} {High-dimensional one-way quantum processing implemented on
  d-level cluster states},\ }\href {https://doi.org/10.1038/s41567-018-0347-x}
  {\bibfield  {journal} {\bibinfo  {journal} {Nature Physics}\ }\textbf
  {\bibinfo {volume} {15}},\ \bibinfo {pages} {148} (\bibinfo {year}
  {2019})}\BibitemShut {NoStop}%
\bibitem [{\citenamefont {Srivastav}\ \emph {et~al.}(2022)\citenamefont
  {Srivastav}, \citenamefont {Valencia}, \citenamefont {McCutcheon},
  \citenamefont {Leedumrongwatthanakun}, \citenamefont {Designolle},
  \citenamefont {Uola}, \citenamefont {Brunner},\ and\ \citenamefont
  {Malik}}]{Srivastav2022}%
  \BibitemOpen
  \bibfield  {author} {\bibinfo {author} {\bibfnamefont {V.}~\bibnamefont
  {Srivastav}}, \bibinfo {author} {\bibfnamefont {N.~H.}\ \bibnamefont
  {Valencia}}, \bibinfo {author} {\bibfnamefont {W.}~\bibnamefont
  {McCutcheon}}, \bibinfo {author} {\bibfnamefont {S.}~\bibnamefont
  {Leedumrongwatthanakun}}, \bibinfo {author} {\bibfnamefont {S.}~\bibnamefont
  {Designolle}}, \bibinfo {author} {\bibfnamefont {R.}~\bibnamefont {Uola}},
  \bibinfo {author} {\bibfnamefont {N.}~\bibnamefont {Brunner}},\ and\ \bibinfo
  {author} {\bibfnamefont {M.}~\bibnamefont {Malik}},\ }\bibfield  {title}
  {\bibinfo {title} {Quick quantum steering: Overcoming loss and noise with
  qudits},\ }\href {https://doi.org/10.1103/PhysRevX.12.041023} {\bibfield
  {journal} {\bibinfo  {journal} {Phys. Rev. X}\ }\textbf {\bibinfo {volume}
  {12}},\ \bibinfo {pages} {041023} (\bibinfo {year} {2022})}\BibitemShut
  {NoStop}%
\end{thebibliography}%
	
	\appendix
	\onecolumngrid
	
	\section{Simulating all depolarised quantum states}
	\label{Appendix_simulation}
	
	\subsection{Arbitrary case}
	\label{Arb_ens}
	We give an explicit $r$-dimensional quantum simulation model for any ensemble ${\cal E}$ of arbitrary cardinality, $m$, comprised of pure states, $\ket{\psi_x}\in\mathbb{C}^d$, subject to depolarising noise,
	\begin{equation}
		\rho_x= v\ketbra{\psi_x}+\frac{1-v}{d} \openone_d,
	\end{equation}
	for some visibility $v\in [0,1]$.
	The protocol to build the simulation is as follows:
	\begin{enumerate}
		\item We choose an arbitrary orthonormal basis of $\mathbb{C}^d$. To fix ideas,  we consider the computational basis $\{\ket{1},\ket{2},\dots,\ket{d}\}$.
		\item We perform the same $d\times d$ unitary transformation $U$ to each of the elements in the basis and take the $r$ first elements of this new set. 
		\item These $r$ states span an $r$-dimensional space. We denote the projector onto this space by
		\begin{equation}
			\Pi^{(r)}_U=\sum_{i=1}^r U\ketbra{i} U^\dagger,
		\end{equation}
		where in the notation we have made explicit its dependence on the unitary $U$.
		\item We simulate each $\rho_x$ by averaging over the Haar measure the normalized projection $\Pi^{(r)}_U\ket{\psi_x}$.
	\end{enumerate} 
	
	Hence, we are claiming that
	\begin{equation}\label{Eq:QSim}
		\rho_x\stackrel{!}{=}\int d\mu_{\text{Haar}}\left(U\right)\frac{\Pi^{(r)}_U\ketbra{\psi_x}\Pi^{(r)}_U}{\bra{\psi_x}\Pi^{(r)}_U\ket{\psi_x}}.
	\end{equation}
	The integral on the right-hand side is invariant under any unitary transformation $U_x$ that leaves invariant the state $\ket{\psi_x}$. Namely,  using the definition of $\Pi^{(r)}_U$, the invariance of $\ket{\psi_x}$ under $U_x$ and the properties of the Haar measure (in particular left and right invariance):
	\begin{equation}
		\begin{aligned}
			&\int d\mu_{\text{Haar}}\left(U\right)U_x\frac{\Pi^{(r)}_U\ketbra{\psi_x}\Pi^{(r)}_U}{\bra{\psi_x}\Pi^{(r)}_U\ket{\psi_x}}U_x^\dagger\\
			&=\int d\mu_{\text{Haar}}\left(U\right)\frac{\Pi^{(r)}_{U_x\, U}\,U_x\ketbra{\psi_x}U_x^\dagger\,\Pi^{(r)}_{U_x\, U}}{\bra{\psi_x}U_x^\dagger\,\Pi^{(r)}_{U_x\, U}\,U_x\ket{\psi_x}}\\
			&=\int d\mu_{\text{Haar}}\left(U\right)\frac{\Pi^{(r)}_{U_x\, U} \ketbra{\psi_x}\Pi^{(r)}_{U_x\, U}}{\bra{\psi_x}\Pi^{(r)}_{U_x\, U} \ket{\psi_x}}\\
			&=\int d\mu_{\text{Haar}}\left(U\right)\frac{\Pi^{(r)}_U \ketbra{\psi_x}\Pi^{(r)}_U}{\bra{\psi_x}\Pi^{(r)}_U \ket{\psi_x}}.
		\end{aligned}
	\end{equation}
	Since the only states invariant under these kinds of transformations (see Appendix \ref{Appendix_invariance_isotropic}) are of the form
	\begin{equation}
		v\ketbra{\psi_x}+\frac{1-v}{d} \openone_d
	\end{equation}
	for some visibility $v\in [0,1]$, we must have that the simulation actually gives a state of this specific form. In order to compute the associated visibility, we take the expected value over $\ket{\psi_x}$:
	\begin{equation}
		\frac{(d-1)v+1}{d}=\int d\mu_{\text{Haar}}\left(U\right)\bra{\psi_x}\Pi^{(r)}_U\ket{\psi_x}.
	\end{equation}
	Expanding the right-hand side we obtain the following sum:
	\begin{equation}
		\sum_{i=1}^{r}\int d\mu_{\text{Haar}}\left(U\right)\abs{\bra{\psi_x}U \ket{i}}^2.
	\end{equation}
	Due to the properties of the Haar measure, each integral in the sum gives the same result:
	\begin{equation}
		\sum_{i=1}^{r}\int d\mu_{\text{Haar}}\left(U\right)\abs{\bra{\psi_x}U \ket{i}}^2=r \int d\mu_{\text{Haar}}\left(\psi\right)\abs{\bra{\psi}\ket{1}}^2,
	\end{equation}
	where we have further identified the integration over unitary transformations with that over pure states.
	
	This last integral can be computed applying the techniques developed in \cite{Jones_2007}. For instance,  given an orthonormal basis $\{\ket{\phi_j}\}_{j=1}^d$ we parametrize any non-normalized pure state $\tilde{\ket{\psi}}$ by 
	\begin{equation}
		\tilde{\ket{\psi}}=\frac{1}{\sqrt{d}}\sum_{j=1}^d z_j \ket{\phi_j},
	\end{equation}
	where $z_j$ are zero-mean Gaussian random variables with the properties $\langle z_j^\ast z_k \rangle = \delta_{j,k}$ and $\langle z_j z_k\rangle =0$. Writing $\tilde{\ket{\psi}}= m \ket{\psi}$, we denote the measure over this ensemble as $d\mu_G\left(\psi,m\right)$.  It can be seen \cite{Jones_2007} that the measure factorizes as
	\begin{equation}
		d\mu_G\left(\psi,m\right)=d\mu_{\text{Haar}}\left(\psi\right)d\mu_G\left(m\right), \quad \int d\mu_G\left(m\right) m^2=1.
	\end{equation}
	For simplicity,  we define $z_j=\sqrt{u_j} e^{i \theta_j}$ so that 
	\begin{equation}
		d\mu_G\left(\psi,m\right)=\frac{1}{(2\pi)^d}\exp{-\sum_{j=1}^d u_j} du_1\cdots du_d d\theta_1\cdots d\theta_d.
	\end{equation}
	
	In addition, since $\{\ket{\phi_j}\}_{j=1}^d$ is arbitrary we can take $\ket{\phi_1}=\ket{1}$. Hence,  $\abs{\bra{\tilde \psi}\ket{1}}^2=\frac{u_1}{d}$ and the integration over the phases is trivial (see Appendix B.3 of \cite{Jones_2007} for more details). The final integral reads
	\begin{equation}
		\begin{aligned}
			&\int d\mu_{\text{Haar}}\left(\psi\right)\abs{\bra{\psi}\ket{1}}^2=\int d\mu_{G}\left(\psi,m\right)\abs{\bra{\tilde \psi}\ket{1}}^2 \\ 
			&= \frac{1}{d}\int_{0}^{\infty} du_1 u_1\int_{0}^{\infty} du_2\cdots\int_{0}^{\infty} du_d\exp{-\sum_{j=1}^d u_j} \\
			&=\frac{1}{d}.
		\end{aligned}
	\end{equation}
	Finally, 
	\begin{equation}
		\begin{aligned}
			\frac{(d-1)v+1}{d}=\frac{r}{d} \quad \implies \quad v=\frac{r-1}{d-1}.
		\end{aligned}
	\end{equation}
	
	\subsection{$s< d$ case}
	\label{Appendix_s_smaller_d-case}
	
	The previous simulation is valid for any noisy ensemble. However, it can be further improved when the initial $m$ pure states span an $s$-dimensional space with $s< d$.  We distinguish in addition the cases $s<m$ and $s=m$.  For the former,  the whole physical $d$-dimensional space is spanned by the ensemble of initial pure states and a complementary ensemble of $d-s$ orthonormal states.  The simulation of the noisy ensemble is then performed by the convex sum of two states:
	\begin{equation}
		\rho_x\stackrel{!}{=} \alpha \rho_x^{(1)}+ (1-\alpha)\frac{\openone_{d-s}}{d-s}
	\end{equation}
	The second state is always $r$-simulable for any $r\geq1$.
	In order to simulate the first state, $\rho_x^{(1)}$, we follow a protocol analogous to the one given in Appendix \ref{Arb_ens} but considering $s\leftrightarrow d$ and $r\leq s$:
	\begin{enumerate}
		\item We choose an arbitrary orthonormal basis of the space $\mathbb{C}^s$ spanned by the ensemble. 
		\item We perform the same $s\times s$ unitary transformation $U$ to each of the elements in the basis and take the $r$ first elements of this new set. 
		\item These $r$ states span an $r$-dimensional space. We denote the projector onto this space by $\Pi^{(r)}_U$, where in the notation we have made explicit its dependence on the unitary $U$.
		\item We simulate the first state in the convex sum by averaging over the Haar measure of the normalized projection $\Pi^{(r)}_U\ket{\psi_x}$.
	\end{enumerate} 
	Thus, we are claiming that
	\begin{equation}\label{Eq:Qs1Sim}
		\rho_x^{(1)}=\int d\mu_{\text{Haar}}\left(U\right)\frac{\Pi^{(r)}_U\ketbra{\psi_x}\Pi^{(r)}_U}{\bra{\psi_x}\Pi^{(r)}_U\ket{\psi_x}}=\frac{r-1}{s-1}\ketbra{\psi_x}+\frac{s-r}{s-1} \frac{\openone_s}{s}.
	\end{equation}
	Hence, solving the system for $(\alpha, v)$ that arises from the matching
	\begin{equation}
		\alpha \rho_x^{(1)}+(1-\alpha)\frac{\openone_{d-s}}{d-s}=v\ketbra{\psi_x} +(1-v)\frac{\openone_d}{d},
	\end{equation}
	leads to 
	\begin{equation}
		\left.
		\begin{aligned}
			v&=&\alpha\frac{r-1}{s-1}\quad \\
			\frac{1-v}{d}&=&\frac{1-\alpha}{d-s}\quad
		\end{aligned}
		\right\}\implies\left\{
		\begin{aligned}
			&\alpha=\frac{s-1}{d-1-r(d/s-1)}\\
			&v=\frac{r-1}{d-1-r(d/s-1)}=\frac{r-1}{d-1}\left(1+\frac{r(d-s)}{d(s-r)+s(r-1)}\right)\geq \frac{r-1}{d-1}
		\end{aligned}\right.
	\end{equation}
	
	For $s=m$,  the whole physical $d$-dimensional space is spanned by the ensemble of initial pure states and a complementary ensemble of $d-m$ orthonormal states.  The simulation of the noisy ensemble is then performed by the convex sum of two states:
	\begin{equation}
		\rho_x\stackrel{!}{=} \alpha \rho_x^{(1)}+ (1-\alpha)\rho_x^{(2)}
	\end{equation}
	
	In order to simulate the first state, $\rho_x^{(1)}$, we follow a protocol analogous to the previous one but considering $m\leftrightarrow d$ and $r\leq m$:
	\begin{enumerate}
		\item We choose an arbitrary orthonormal basis of the space $\mathbb{C}^m$ spanned by the ensemble. 
		\item We perform the same $m\times m$ unitary transformation $U$ to each of the elements in the basis and take the $r$ first elements of this new set. 
		\item These $r$ states span an $r$-dimensional space. We denote the projector onto this space by $\Pi^{(r)}_U$, where in the notation we have made explicit its dependence on the unitary $U$.
		\item We simulate the first state in the convex sum by averaging over the Haar measure of the normalized projection $\Pi^{(r)}_U\ket{\psi_x}$.
	\end{enumerate} 
	Thus, we are claiming that
	\begin{equation}\label{Eq:Qm1Sim}
		\rho_x^{(1)}=\int d\mu_{\text{Haar}}\left(U\right)\frac{\Pi^{(r)}_U\ketbra{\psi_x}\Pi^{(r)}_U}{\bra{\psi_x}\Pi^{(r)}_U\ket{\psi_x}}=\frac{r-1}{m-1}\ketbra{\psi_x}+\frac{m-r}{m-1} \frac{\openone_m}{m}.
	\end{equation}
	Concerning the second state entering the convex sum, $\rho_x^{(2)}$, we proceed as follows:
	\begin{enumerate}
		\item We take an arbitrary orthonormal basis for the complementary $(d-m)$-dimensional space, $\{\ket{\phi_y}\}_{y=m+1}^d$.  For each element in this basis,  $\ket{\phi_y}$, we build all possible projectors onto the spaces spanned by this state and $r-1$ elements from $\{\ket{\psi_x}\}_{x=1}^m$. We then have in total $(d-m){m \choose r-1}$ $\Pi_\lambda$ projectors .
		\item The state $\sigma_{x,\lambda}$ in the simulation is given by $\ket{\psi_x}$ if it belongs to the space in hand and $\ket{\phi_y}$ otherwise.  In addition, we take a uniform probability distribution $q_\lambda\equiv \left((d-m){m \choose r-1}\right)^{-1}$.  
	\end{enumerate}
	Thus, we are claiming that 
	\begin{equation}\label{Eq:Qm2Sim}
		\rho_x^{(2)}=\frac{1}{(d-m){m \choose r-1}}\left((d-m){m-1 \choose r-2} \ketbra{\psi_x} +  {m-1 \choose r-1}\openone_{d-m} \right)=\frac{r-1}{m} \ketbra{\psi_x} +  \frac{m-(r-1)}{m}\frac{\openone_{d-m}}{d-m}.
	\end{equation}
	Therefore, taking $\alpha=\frac{(m-1)(m-r+1)}{d(m-r)+(r-1)}$:
	\begin{equation}
		\alpha \rho_x^{(1)}+ (1-\alpha)\rho_x^{(2)}=\frac{(r-1)\left(d(m-r)+m\right)}{m\left(d(m-r)+(r-1)\right)}\ketbra{\psi_x} +\frac{d(m-r)(m-(r-1))}{m\left(d(m-r)+(r-1)\right)}\frac{\openone_d}{d},
	\end{equation}
	that corresponds to a noisy state with visibility 
	\begin{equation}
		v=\frac{(r-1)\left(d(m-r)+m\right)}{m\left(d(m-r)+(r-1)\right)}=\frac{r-1}{d-1-r(d/m-1)}\left(1+(d-m)\frac{d}{m}\frac{m-r}{m}\frac{m-r}{d(m-r)+r-1}\right)\geq\frac{r-1}{d-1-r(d/m-1)}.
	\end{equation}

	\newpage
	\section{Quantum simulation orthonormal ensemble}
	\label{Appendix_finite-sim}
	Let us consider that ${\cal E}=\{\rho_x\}_{x=1}^d$ is comprised of $d$ orthonormal pure states, $\ket{x}\in\mathbb{C}^d$, subject to depolarising noise:
	\begin{equation}
		\rho_x=v\ketbra{x} +\frac{1-v}{d}\openone,
	\end{equation}
	for some visibility $v\in[0,1]$. For this ensemble, the $r$-dimensional simulation leading to $v_{\text{crit}}$ is achieved using finitely many subspaces.
	
	The protocol to build the simulation is as follows:
	\begin{enumerate}
		\item We build all possible $r$-dimensional projectors using the elements of the basis, i.e.  all possible $d \choose  r$ projectors $\Pi_\lambda$, and take a uniform distribution over them, $q\left(\lambda\right)\equiv{d\choose r}^{-1}$ .
		\item For each state in the ensemble, $\rho_x$ and each projector onto a subspace, $\Pi_\lambda$: if $\bra{x}\Pi_\lambda\ket{x}=1$ we take $\ketbra{x}$ as the state to be used for the simulation. If not, and hence $\bra{x}\Pi_\lambda\ket{x}=0$, we take the normalised $\Pi_\lambda/r$ projector as the state to be used for the simulation.
	\end{enumerate} 
	For instance,  the number of projectors containing a fixed pure state $\ket{x}$ is $d-1 \choose r-1$.  Besides, the sum over all other projectors yields to ${d-2 \choose r-1}\frac{1}{r}\left(\openone-\ketbra{x}\right)$. Hence,  the simulation gives
	\begin{equation}
		\dfrac{1}{{d \choose r}}\left[{d-1 \choose r-1}\ketbra{x}+{d-2 \choose r-1}\frac{1}{r}\left(\openone-\ketbra{x}\right)\right].
	\end{equation}
	Simplifying the above expression we get
	\begin{equation}
		\frac{r-1}{d-1}\ketbra{x}+\frac{d-r}{d-1}\frac{1}{d}\openone,
	\end{equation}
	which coincides with the noisy state associated with the visibility $v_{\text{crit}}$.
	
	When the ensemble is comprised of a subset of orthonormal pure states with cardinality $m\leq d$ and $m\geq r$,  the visibility $v=\frac{(r-1)\left(d(m-r)+m\right)}{m\left(d(m-r)+(r-1)\right)}$ can be obtained using as well a finite simulation. This is clear since the protocol to get this visibility is divided into two simulations: the one for the state $\rho_x^{(1)}$ based on infinitely many subspaces and the one for $\rho_x^{(2)}$ which already uses finitely many subspaces.  Nevertheless, we have just shown that the simulation for $\rho_x^{(1)}$ can be achieved using a finite number of subspaces,  and in conclusion so is the case for the total simulation.

	\section{Invariance of states mixed with white noise}
	\label{Appendix_invariance_isotropic}
	In this appendix we show that the family of states $\rho$ such that
	\begin{equation}
		\forall U_x\in {\cal U}(d) :\ U_x\ket{\psi_x}=\ket{\psi_x},\quad U_x\, \rho \,U_x^\dagger=\rho,
	\end{equation}
	has to be of the form
	\begin{equation}
		v\ketbra{\psi_x}+\frac{(1-v)}{d}\openone_d,
	\end{equation}
	for some visibility $v\in[0,1]$. 
	
	In the first place, we notice that under these assumptions $\ket{\psi_x}$ is an eigenvector of $\rho$. To see this, let $\ket{\varphi}\equiv\rho \ket{\psi_x}$.  Due to the invariance of both $\rho$ and $ \ket{\psi_x}$ under $U_x$, it holds that:
	\begin{equation}
		U_x\ket{\varphi}=U_x\, \rho \ket{\psi_x}=\rho\, U_x  \ket{\psi_x}=\rho  \ket{\psi_x}=\ket{\varphi}.
	\end{equation}
	Hence, for any $ U_x\in {\cal U}(d)$ such that $U_x\ket{\psi_x}=\ket{\psi_x}$, we have that $U_x\ket{\varphi}=\ket{\varphi}$. This implies that $\rho \ket{\psi_x}=\ket{\varphi}\propto \ket{\psi_x}$, i.e.  $\ket{\psi_x}$ is an eigenvector of $\rho$.  We denote the eigenvalue associated with $\ket{\psi_x}$ by $\tilde \lambda$.
	
	Denoting by $\ket{\varphi_i}$ and $\lambda_i$, for $i=1,\dots,d-1$, the other $d-1$ orthonormal eigenvectors and eigenvalues, we know that
	\begin{equation}
		\rho=\tilde \lambda \ketbra{\psi_x} +\sum_{i=1}^{d-1} \lambda_i \ketbra{\varphi_i}.
	\end{equation}
	In the basis $\{\ket{\psi_x},\ket{\varphi_i}\}$,  the matrix $U_x$ takes the form
	\begin{equation}
		U_x=1\oplus U_x^{(d-1)},\quad U_x^{(d-1)}\in {\cal U}(d-1).
	\end{equation}
	Because the condition $U_x\, \rho \,U_x^\dagger=\rho$ must hold for any unitary, we can take as a particular case those for which $U_x^{(d-1)}$ permutes the basis elements $\ket{\varphi_1}$ and $\ket{\varphi_i}$, with $i=2,\dots,d-1$. For that case, we have:
	\begin{equation}
		\begin{aligned}
			0=\rho-U_x\, \rho \,U_x^\dagger=&(\lambda_1-\lambda_i) \ketbra{\varphi_1}+(\lambda_i-\lambda_1) \ketbra{\varphi_i}\\
			&\implies \lambda_1=\lambda_i.
		\end{aligned}
	\end{equation}
	Since the choice of $i$ was arbitrary, we then deduce that $\forall  i\ \lambda_i=\lambda_1=\lambda$. Thus, 
	\begin{equation}
		\rho=\tilde \lambda \ketbra{\psi_x} + \lambda \sum_{i=1}^{d-1} \ketbra{\varphi_i}=(\tilde \lambda-\lambda) \ketbra{\psi_x}+\lambda\openone_d.
	\end{equation}
	Redefining the combination $\tilde \lambda-\lambda$ as the visibility $v$ and imposing that $\rho$ is a positive semi-definite matrix with $\Tr{\rho}=1$, we finally get:
	\begin{equation}
		\rho=v\ketbra{\psi_x}+\frac{(1-v)}{d}\openone_d,
	\end{equation}
	for some visibility $v\in[0,1]$.

\end{document}